\documentclass[letterpaper, 10 pt, conference]{ieeeconf}  

\IEEEoverridecommandlockouts                              

\usepackage[font=small,labelfont=bf]{caption}
\usepackage{graphics} 
\usepackage{epsfig} 
\usepackage{mathptmx} 
\usepackage{times} 
\usepackage{amsmath} 
\usepackage{amssymb}  
\usepackage{caption}
\usepackage{subcaption}
\title{\LARGE \bf
Deep Learning based Skin-layer Segmentation for Characterizing Cutaneous Wounds from Optical Coherence Tomography Images
}

\author{Prashant Kumar, Swatantra Dhara, Ayan Gope, Jyotirmoy Chatterjee, Subhamoy Mandal$^{\star}$, \textit{Senior Member, IEEE}}

\begin{document}

\maketitle
\thispagestyle{empty}
\pagestyle{empty}

\begin{abstract}

Optical coherence tomography (OCT) is a medical imaging modality that allows us to probe deeper sub-structures of skin. The state-of-the-art wound care prediction and monitoring methods are based on visual evaluation and focus on surface information. However, research studies have shown that sub-surface information of the wound is critical for understanding the wound healing progression. This work demonstrated the use of OCT as an effective imaging tool for objective and non-invasive assessments of wound severity, the potential for healing, and healing progress by measuring the optical characteristics of skin components. We have demonstrated the efficacy of OCT in studying wound healing progress \textit{in vivo} small animal models. Automated analysis of OCT datasets poses multiple challenges, such as limitations in the training dataset size, variation in data distribution induced by uncertainties in sample quality and experiment conditions. We have employed a U-Net-based model for segmentation of skin layers based on OCT images and to study epithelial and regenerated tissue thickness wound closure dynamics and thus quantify the progression of wound healing. In the experimental evaluation of the OCT skin image datasets, we achieved the objective of skin layer segmentation with an average intersection over union (IOU) of 0.9234. The results have been corroborated using gold-standard histology images and co-validated using inputs from pathologists.

\indent \textit{Clinical Relevance}\textemdash To monitor wound healing progression without disrupting the healing procedure by superficial, non-invasive means via the identification of pixel characteristics of individual layers.
\end{abstract}

\let\thefootnote\relax\footnotesize\footnotetext{
P Kumar and S Mandal are with the School of Medical Science and Technology, Indian Institute of Technology Kharagpur, Kharagpur, India, PIN-721302. (e-mail: smandal@smst.iitkgp.ac.in)\\
J Chatterjee is with the Dr B C Roy Multi Speciality Medical Research Centre, Indian Institute of Technology Kharagpur, Kharagpur, India\\
S Dhara is with the Center for Health Science and Technology, JIS Institute of Advanced Studies and Research, Kolkata, India\\
A Gope is with the Advanced Technology Development Centre, Indian Institute of Technology, Kharagpur, India
}

\section{INTRODUCTION}
The wound healing process is a complex procedure that involves four distinct stages: homeostasis, inflammation, proliferation, and remodeling. This process involves a variety of cells, growth factors, and cellular and molecular mechanisms, and its pathophysiology is still being studied to optimize treatment and improve patient outcomes. Chronic wounds, such as diabetic and venous ulcers, have stopped healing normally due to certain pathophysiologic conditions. The gold standard for precise tissue architecture assessment is still histological analysis, but this process is invasive and can cause additional trauma. Most wound care prediction and monitoring currently relies on visual evaluation, leaving a critical need for novel, non-invasive strategies for individualized diagnosis and care\cite{wound}. Imaging technologies can provide quick, objective, and non-invasive assessments of wound severity, potential for healing, and healing progress by measuring the optical characteristics of skin components. Various imaging modalities have been used to study cutaneous wounds, including near-infrared imaging, thermal imaging, optical coherence tomography (OCT), fluorescence imaging, and laser speckle imaging. OCT in particular provides high-resolution images of tissue microstructure and has been used to assess burn wound severity, monitor scar progression, and investigate chronic wounds. Since OCT can accurately distinguish between the epidermal and dermal layers, it can be utilized in clinical settings to track wound re-epithelization\cite{gope}. To establish OCT as a standalone tool for non-invasive wound evaluation, it is necessary to correlate its data with corresponding histological data\cite{OCT} and validate it with a large sample set of optical data. The present pilot study aims to use OCT as a precise and accurate tool for non-invasive wound evaluation by developing a Convolutional Neural Network (CNN) based model that can automatically segment OCT images to annotate different layers of normal and wounded skin tissue and help quantitatively analyze the healing stage, epithelial thickness, regenerated tissue thickness, and wound closure dynamics. Previous research has focused on the use of deep learning and other techniques for the segmentation of the epidermis layer in OCT images, which is useful for analyzing wound epithelial reformation\cite{3},\cite{5},\cite{6},\cite{7},\cite{8},\cite{9},\cite{10}. However, segmenting deeper layers such as the dermis and subcutaneous tissue is equally important to assess the dynamics of wound healing\cite{11} fully,\cite{12},\cite{debdoot}. Our study aims to fill this gap by developing a U-Net-based approach for segmenting three layers (epidermis, dermis, and subcutaneous) of wounded skin in OCT images. We selected the U-Net architecture because of its effectiveness in biomedical image segmentation\cite{Unet_paper}.\\
In our work, we tackled the difficulties posed by autonomous systems, particularly in optical coherence tomography datasets, such as limited training data and image data distribution dependency. To overcome these obstacles, we utilized transfer learning, a common approach in the field. Using segmentation models with pre-trained models, we could apply Machine Learning (ML) models in situations where large, well-curated training datasets are difficult to obtain. In the case of optical coherence tomography, where creating high-quality annotated image datasets is both costly and technically challenging, reusing pre-trained models helps to implement ML models more efficiently. Our study utilized the base U-net and used pre-trained models such as VGG16, ResNet34, and InceptionV3.\newline
Transfer learning helps address the problem of limited data, but uncertainty quantification is computationally expensive with single models, leading to an ensemble of pre-trained models. Ensemble learning is a method in artificial intelligence (AI)\cite{22} where multiple models or experts are used to solve an AI problem. Suppose we have k models in an ensemble, denoted as $(m_{1}, . . . , m_{k})$, and let $ p(x = y_{i}|m_{j})$ denote the probability that input x is classified as $y_{i}$ under model $m_{j}$. Then, the ensemble prediction is given by:
\begin{equation}
    p(x = y_{i}|m_{1}, . . . , m_{k}) =\frac{1}{k}\sum_{j=1}^{k} p(x = y_{i}| m_j)
\end{equation}
The goal is to increase diversity among the models, accurate segmentation, good calibration, and reliable uncertainty quantification, and avoid overfitting. The final prediction is obtained by combining the outputs of the individual models. We also compared the performance of multiple pre-trained models as the encoder part of the U-Net. We evaluated the potential improvement obtained through the ensemble of those models. We calculated the wound bed area from the segmented output. The details of the ensemble design, transfer learning, and training dataset preparation are outlined in Section 2, and the significant results and discussion are provided in Section 3. 


\begin{figure*}[htb!]
    \begin{center}
            \begin{subfigure}[b]{0.32\textwidth}
                \centering
                \includegraphics[width=0.95\linewidth]{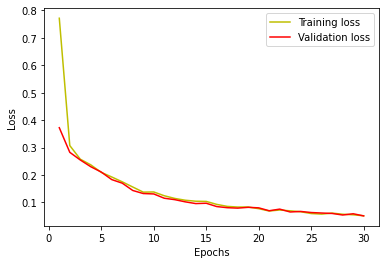}
                \caption{Base U-net loss curve}
                \label{fig:gull}
            \end{subfigure}%
            \begin{subfigure}[b]{0.32\textwidth}
                \centering
                \includegraphics[width=0.95\linewidth]{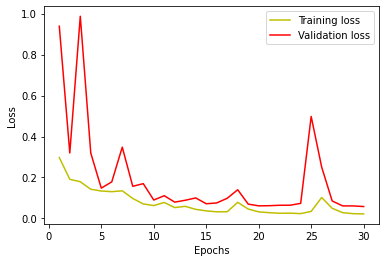}
                \caption{ResNet34 U-net architecture loss curve}
                \label{fig:gull2}
            \end{subfigure}%
            \begin{subfigure}[b]{0.32\textwidth}
                \centering
                \includegraphics[width=0.95\linewidth]{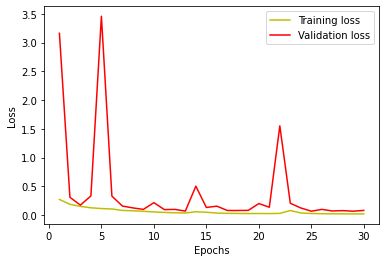}
                \caption{Inceptionv3 U-net architecture loss curve}
                \label{fig:tiger}
            \end{subfigure}%
        \caption{\textbf{Training and validation loss for different U-net architectures.}}\label{fig:loss}
    \end{center} 
\end{figure*}

\section{Materials and Methods}
\subsection{Data acquisition}
The \textit{in vivo}  experiments were carried out on a mice wound model according to guidelines approved by Institutional Animal Ethics Committee (IAEC Authorization- IE-2/JC-SMST/2.18, sanctioned by IIT Kharagpur), following the guiding principles set by the Committee for The Purpose of Control and Supervision On Experiments On Animals (CPCSEA) in accordance with National Research Council's Guide for the Care and Use of Laboratory Animals. In this study, full-thickness cuts were inflicted on the skin of albino mice, and the wounds were allowed to heal without intervention. The healing process was monitored by acquiring several wound images taken from different tissue sections at different time points (0, 4, 8, and 12 days) using SS-OCT (Model: OCS1300SS, manufactured by Thorlabs-Inc.). The SS-OCT instrument had a center wavelength of 1325 nm, a half-power spectral bandwidth of less than 100 nm, an axial scan rate of 16 kHz, a coherence length of 6.0 nm, and an average output power of 10 mW. In total, 82 images were taken from two animals across the study period, with 7 images having to be disregarded due to poor image quality.
\subsection{OCT datasets labeling}
\label{sec: OCT datasets labeling}

Three layers were labeled by an expert. Before labeling all images were despeckled and applied with a median filter for better interpretability of the layers. Epidermis, dermis, and subcutaneous layers were labeled with mask values of 1, 2, and 3. The labeling was done with the help of the Labkit toolbox of ImageJ. A senior expert supervised the labeling done by the first expert.
To assess the reliability of the image annotations, a small subset of the images was independently annotated by a medical professional and the inter-annotator agreement was measured using Cohen's kappa score. The resulting score of 0.74 indicates a substantial level of agreement between the annotators.
\begin{figure}[hb!]
    \centering
    \includegraphics[width=1\linewidth]{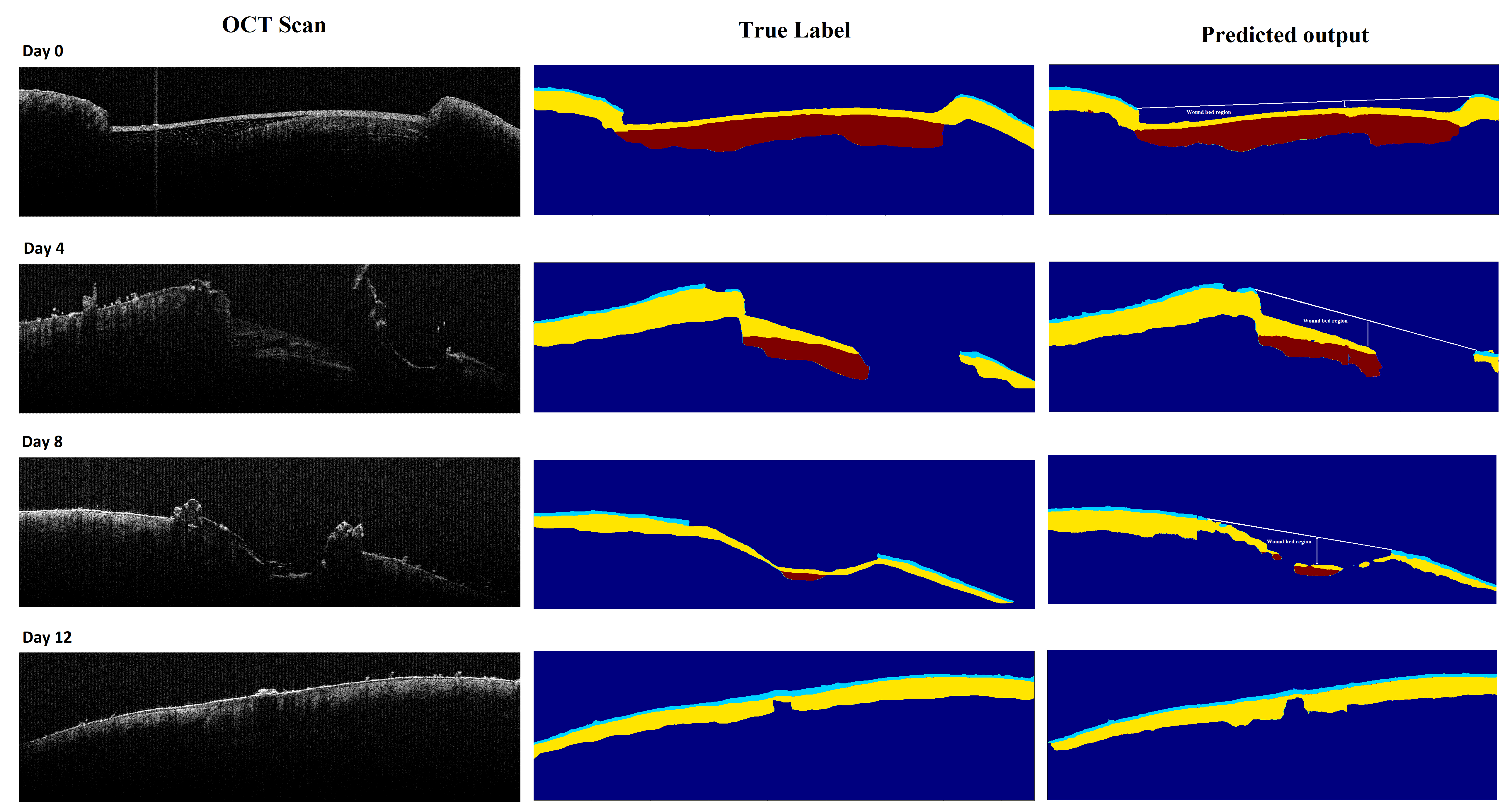}
    \caption{\textbf{Segmentation results at different stages of healing.}
Column 1 displays raw OCT images for all days. Skin layers are represented in other images in color code: Cyan: Epidermis, Yellow: Dermis, Red: Subcutaneous.
}
    \label{fig:segmented}
\end{figure}
\subsection{Methodology}

Biomedical image segmentation is characterized by several specific factors and it is a difficult task to process cross-sectional OCT images with hyper-reflection, bulk motion, and limited training data. Thus, finding proper architecture is an essential step in our study. To overcome the limitation of small datasets the 75 images were broken down into 690 patches. It is very common in OCT image processing that the anatomy of interest only occupies a very small portion of the image. Hence, most of the extracted patches belong to the background area, often resulting in the trained network being biased toward the background and getting trapped in local minima\cite{15}. To mitigate the issue, we filtered out patches containing only background by applying a thresholding operation on the generated patches used for training. Patches with a unique value greater than or equal to a defined threshold in their mask were considered for training, resulting in a final training dataset size of 660.
In our study, we implemented the base U-net and utilized pre-trained models VGG16\cite{18}, ResNet34\cite{17}, and InceptionV3\cite{19}. To enhance the quality of segmentation, we employed ensemble techniques on multiple pre-trained U-Net models and calculated individual weights for each architecture. The optimal weight combination for the ResNet34, InceptionV3, and VGG16 models were determined to be 0.3, 0.2, and 0.5 respectively, through automatic optimization. The data was split into 80\% for training and 20\% for validation during the model training process. The training and validation process utilized a learning rate of $10^{-5}$, involved 30 training epochs with the Adam optimizer, and used the categorical-cross-entropy loss as the evaluation metric with Intersection over Union (IoU).
The best model parameters were saved after training and validation and the loss comparison between training and validation is depicted in Figure \ref{fig:loss}.

\begin{table*}
    \centering
    \begin{tabular}{c|c|c|c|c|c}
        \hline
        \textbf{U-Net Architecture} & \textbf{IoU (Background)} & \textbf{IoU (Epidermis)} & \textbf{IoU (Dermis)} & \textbf{IoU (Subcutaneous)} & \textbf{Mean IoU} \\
        \hline
        Base U-Net & 0.9881 & 0.8186  & 0.8897 & 0.9094 & 0.9014 \\
        ResNet34 & 0.9832 & 0.8312 & 0.9135 & 0.9152 & 0.9108 \\
        InceptionV3 & 0.9808 & 0.8461 & 0.9092 & 0.8924 & 0.9071 \\
        VGG16 & 0.9758 & 0.8411 & 0.8949 & 0.8615 & 0.8933\\
        Ensemble Pretrained Architecture & -  & - & - & - & 0.9234 \\
        
        \hline
    \end{tabular}
    \caption{\textbf {IoU Score of different U-net architecture and segmentation performance.}}
    \label{tab:results}
\end{table*}


\begin{figure}[htb!]
    \centering
    \includegraphics[width=0.95\linewidth]{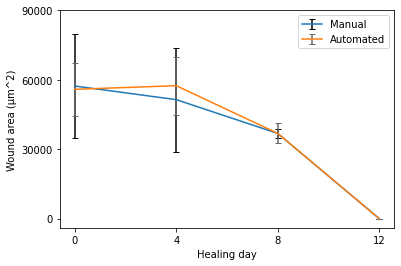}
    \caption{\textbf{Wound healing progression graph.}}
    \label{fig:method}
\end{figure}

\begin{figure}[ht!]
    \centering
    \includegraphics[width=0.95\linewidth]{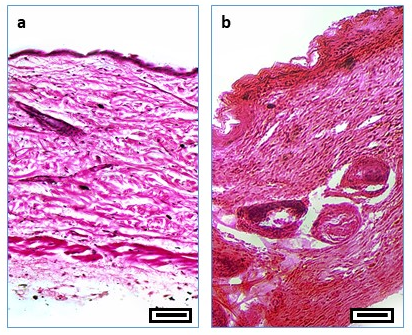}
    \caption{\textbf{Histopathology validation} (Magnification = 100X and scale bar = 10 µm for all images): “a” shows HE-stained normal, healthy skin while “b” shows HE-stained section of a healed wound region. Differences in the thickness of the keratinized layer and the density of collagen in the granulation region are a few of the characteristics clearly visible from these sections.}
    \label{fig:histopathology}
\end{figure}

\section{Result and Discussion}
The results as demonstrated in Table \ref{tab:results} show that all models successfully segmented the different layers in the wound images. Figure \ref{fig:segmented} displays the progression of wound healing over time, demonstrating the ability of the models to accurately track the healing process, saving time and effort. This also allowed for the calculation of wound bed area and the development of a wound healing progression function (Figure \ref{fig:method}). The use of a patch-based augmentation technique and small batch size resulted in a fast training time, as evidenced by the training graphs (\ref{fig:loss}). Although all models performed similarly with minor differences in per-class segmentation. However, the ensemble model emerged as the best performer, combining the best results from all models. The wound healing progress (Figure \ref{fig:segmented}) shows that the healing process is not linear but shows only slight changes during day 0 to day 4, and the healing process accelerates thereafter. This demonstrates the critical need of proper wound curation during the initial day post injury. The healing process from day 4 to day 12 (last imaging trial) shows rapid improvement and more than 98\% wound are healed by day 12, unless there are specific aberrations or infection scenarios. However, such cases are beyond the scope of the current study. These results were further corroborated by using histopathology samples (excised wound region). In Figure \ref{fig:histopathology} we show the differences between normal and healed wound regions using Hematoxylin and Eosin (HE) stained histopathology section. The histopathology studies clearly highlights the differences in the thickness of keratinized layer and the density  of collagen in the granulation region. These features can be clearly be seen in the microscopic images and further characterized using image analysis methods to strengthen our understanding of the wound healing process \cite{5712739, rsob.200223}.    

\section{Conclusion}
\label{sec:foot}
In summary, our study demonstrates the effectiveness of deep learning models, specifically U-Net and its variants, in segmenting different skin layers from OCT scans of dorsal cutaneous wounds in mice. We successfully demonstrated the effectiveness of automatic segmentation in calculating the wound bed area to annotate the different stages of wound healing. This is a novel domain of exploration in medical imaging and machine intelligence. The initial sets of experiments have been carried out on two Swiss albino mice, which shows significant feature variations in skin characteristics such as the mean thickness of skin layers, hair density, and concomitantly, high light absorption. However, wounds in mice heal typically by contraction, which gives us a simple mechanism to study for the initial assessment of our tool. We will expand the scope of our studies to include more animals and gradually shift to large animal models which are better candidates for studying wound healing in preclinical phases. In our work, we evaluated various U-Net models and demonstrated the impact of patch-based data augmentation and small batch size on the performance of deep-learning models in medical imaging. Our comparison results indicate that the ensemble model, which combines the best results from multiple models, was the top performer. The findings suggest that deep learning has the potential to be a valuable tool in analyzing wound healing progress \cite{8471151}. Additionally, these results conclusively show the relevance of OCT and similar sub-surface imaging methods in studying wound healing dynamics and changes in tissue characteristics during the healing process. Such investigation can be useful in clinical cases of non-healing and infected wounds. Further, research and joint feature learning from multi-modal and multiscale imaging \cite{ye2019cross, mandal2015multiscale} are needed to improve accuracy and address the limitations \cite{chan2020wound}. Further improvements of relevant techniques can usher in a new era of digital pathology\cite{6099988}, improve access to care in resource-limited settings, and alleviate the clinical and economic burden of wound care in the tropics\cite{wound_burden}. 

\addtolength{\textheight}{-12cm}   





\section*{ACKNOWLEDGMENT}
The authors acknowledge the support of Dr. Shirin Dasgupta,Senior Resident (Pathology) at Dr. B. C. Roy Multi Speciality Medical Research Centre, Indian Institute of Technology, Kharagpur, toward her support for data annotation and study validation.   


\bibliography{refs.bib}

\begin{thebibliography}{10}

\bibitem{wound}
Bodo et~al.,
\newblock ``Multimodal noninvasive monitoring of soft tissue wound healing,''
\newblock {\em Journal of clinical monitoring and computing}, vol. 27, pp.
  677--688, 2013.

\bibitem{gope}
Gope et~al.,
\newblock ``Regenerative repair of full thickness skin wound assisted by dual
  crosslinking percolative gel casting maneuvered alginate hydrogel embedded
  with honey ghee blend resembles standard cutaneous properties,''
\newblock {\em Journal of Tissue Viability}, vol. 31, no. 4, pp. 657--672,
  2022.

\bibitem{OCT}
Barui et~al.,
\newblock ``Swept-source optical coherence tomography of lower limb wound
  healing with histopathological correlation,''
\newblock {\em Journal of biomedical optics}, vol. 16, no. 2, pp.
  026010--026010, 2011.

\bibitem{3}
Yubo~Ji et~al.,
\newblock ``{Deep-learning approach for automated thickness measurement of
  epithelial tissue and scab using optical coherence tomography},''
\newblock {\em Journal of Biomedical Optics}, vol. 27, no. 1, pp. 015002, 2022.

\bibitem{5}
Li~et~al.,
\newblock ``Epidermal segmentation in high-definition optical coherence
  tomography,''
\newblock in {\em 2015 37th Annual International Conference of the IEEE
  Engineering in Medicine and Biology Society (EMBC)}, 2015, pp. 3045--3048.

\bibitem{6}
Jesse Weissman, Tom Hancewicz, and Peter Kaplan,
\newblock ``Optical coherence tomography of skin for measurement of epidermal
  thickness by shapelet-based image analysis,''
\newblock {\em Opt. Express}, vol. 12, no. 23, pp. 5760--5769, Nov 2004.

\bibitem{7}
Srivastava et~al.,
\newblock ``Supervised 3d graph-based automated epidermal thickness
  estimation,''
\newblock in {\em 2017 IEEE 2nd International Conference on Signal and Image
  Processing (ICSIP)}, 2017, pp. 297--301.

\bibitem{8}
G.~Josse, J.~George, and D.~Black,
\newblock ``Automatic measurement of epidermal thickness from optical coherence
  tomography images using a new algorithm,''
\newblock {\em Skin Research and Technology}, vol. 17, no. 3, pp. 314--319,
  2011.

\bibitem{9}
Taghavikhalilbad et~al.,
\newblock ``Semi-automated localization of dermal epidermal junction in optical
  coherence tomography images of skin,''
\newblock {\em Appl. Opt.}, vol. 56, no. 11, pp. 3116--3121, Apr 2017.

\bibitem{10}
Cobb et~al.,
\newblock ``{Noninvasive assessment of cutaneous wound healing using
  ultrahigh-resolution optical coherence tomography},''
\newblock {\em Journal of Biomedical Optics}, vol. 11, no. 6, pp. 064002, 2006.

\bibitem{11}
Kepp et~al.,
\newblock ``Segmentation of mouse skin layers in optical coherence tomography
  image data using deep convolutional neural networks,''
\newblock {\em Biomed. Opt. Express}, vol. 10, no. 7, pp. 3484--3496, Jul 2019.

\bibitem{12}
Gao et~al.,
\newblock ``Automatic segmentation of laser-induced injury oct images based on
  a deep neural network model,''
\newblock {\em International Journal of Molecular Sciences}, vol. 23, no. 19,
  pp. 11079, Sep 2022.

\bibitem{debdoot}
Sheet et~al.,
\newblock ``{In situ histology of mice skin through transfer learning of tissue
  energy interaction in optical coherence tomography},''
\newblock {\em Journal of Biomedical Optics}, vol. 18, no. 9, pp. 090503, 2013.

\bibitem{Unet_paper}
Olaf Ronneberger, Philipp Fischer, and Thomas Brox,
\newblock ``U-net: Convolutional networks for biomedical image segmentation,''
  2015.

\bibitem{22}
Thomas~G Dietterich,
\newblock ``Ensemble methods in machine learning,''
\newblock in {\em Multiple Classifier Systems: First International Workshop,
  MCS 2000 Cagliari, Italy, June 21--23, 2000 Proceedings 1}. Springer, 2000,
  pp. 1--15.

\bibitem{15}
Merkow et~al.,
\newblock ``Dense volume-to-volume vascular boundary detection,''
\newblock in {\em Medical Image Computing and Computer-Assisted Intervention -
  MICCAI 2016}, Cham, 2016, pp. 371--379, Springer International Publishing.

\bibitem{18}
Karen Simonyan and Andrew Zisserman,
\newblock ``Very deep convolutional networks for large-scale image
  recognition,''
\newblock {\em arXiv preprint arXiv:1409.1556}, 2014.

\bibitem{17}
He~et~al.,
\newblock ``Deep residual learning for image recognition,''
\newblock in {\em 2016 IEEE Conference on Computer Vision and Pattern
  Recognition (CVPR)}, 2016, pp. 770--778.

\bibitem{19}
Szegedy et~al.,
\newblock ``Rethinking the inception architecture for computer vision,''
\newblock in {\em Proceedings of the IEEE conference on computer vision and
  pattern recognition}, 2016, pp. 2818--2826.

\bibitem{5712739}
Subhamoy Mandal, Amit Kumar, J~Chatterjee, M~Manjunatha, and Ajoy~K Ray,
\newblock ``Segmentation of blood smear images using normalized cuts for
  detection of malarial parasites,''
\newblock in {\em 2010 Annual IEEE India Conference (INDICON)}, 2010, pp. 1--4.

\bibitem{rsob.200223}
Holly~N. Wilkinson and Matthew~J. Hardman,
\newblock ``Wound healing: cellular mechanisms and pathological outcomes,''
\newblock {\em Open Biology}, vol. 10, no. 9, pp. 200223, 2020.

\bibitem{8471151}
Subhamoy Mandal, Aaron~B. Greenblatt, and Jingzhi An,
\newblock ``Imaging intelligence: Ai is transforming medical imaging across the
  imaging spectrum,''
\newblock {\em IEEE Pulse}, vol. 9, no. 5, pp. 16--24, 2018.

\bibitem{ye2019cross}
Linwei Ye, Mrigank Rochan, Zhi Liu, and Yang Wang,
\newblock ``Cross-modal self-attention network for referring image
  segmentation,''
\newblock in {\em Proceedings of the IEEE/CVF conference on computer vision and
  pattern recognition}, 2019, pp. 10502--10511.

\bibitem{mandal2015multiscale}
Subhamoy Mandal, PS~Viswanath, N~Yeshaswini, X~Lu{\'\i}s Dean-Ben, and Daniel
  Razansky,
\newblock ``Multiscale edge detection and parametric shape modeling for
  boundary delineation in optoacoustic images,''
\newblock in {\em 2015 37th Annual International Conference of the IEEE
  Engineering in Medicine and Biology Society (EMBC)}. IEEE, 2015, pp.
  707--710.

\bibitem{chan2020wound}
Kai~Siang Chan and Zhiwen~Joseph Lo,
\newblock ``Wound assessment, imaging and monitoring systems in diabetic foot
  ulcers: A systematic review,''
\newblock {\em International Wound Journal}, vol. 17, no. 6, pp. 1909--1923,
  2020.

\bibitem{6099988}
Hrushikesh Garud, Ajoy~K. Ray, Subhamoy Mandal, Debdoot Sheet, Manjunatha
  Mahadevappa, and Jyotirmoy Chatterjee,
\newblock ``Volume visualization approach for depth-of-field extension in
  digital pathology,''
\newblock in {\em 2011 4th International Congress on Image and Signal
  Processing}, 2011, vol.~1, pp. 335--339.

\bibitem{wound_burden}
Zhiwen~J. Lo, Xuxin Lim, Diane Eng, Josip Car, Qiantai Hong, Enming Yong,
  Li~Zhang, Sadhana Chandrasekar, Glenn W.~L. Tan, Yam~M. Chan, Seow~C. Sim,
  Chien~W. Oei, Xiaojin Zhang, Ayliana Dharmawan, Yi~Z. Ng, Keith Harding, Zee
  Upton, Chun~W. Yap, and Bee~H. Heng,
\newblock ``Clinical and economic burden of wound care in the tropics: a 5-year
  institutional population health review,''
\newblock {\em International Wound Journal}, vol. 17, no. 3, pp. 790--803,
  2020.

\end{thebibliography}

\bibliographystyle{IEEEbib}
\end{document}